\documentclass[sigconf]{acmart}
\usepackage{xkeyval}
\usepackage{venndiagram}
\usepackage{etoolbox}

\usepackage{enumitem}
\newlist{questions}{enumerate}{2}
\setlist[questions,1]{label=\textbf{RQ}\arabic*.,ref=RQ\arabic*}
\setlist[questions,2]{label=(\alph*),ref=\thequestionsi(\alph*)}

\usepackage[utf8]{inputenc}
\usepackage[T1]{fontenc}
\usepackage{pifont} % Added

\usepackage{amsmath,amsthm}
\usepackage[linesnumbered,ruled,vlined]{algorithm2e}
\usepackage{siunitx} % Scientific units

\usepackage{graphicx}      % Include figures
\graphicspath{{./images/}} % Set default graphics path
\usepackage{caption}       % Sub-figures
\usepackage{subcaption}    % Sub-figures
\usepackage{tikz}          % Tikz graphics
\usetikzlibrary{arrows}    % Tikz graphics
\usetikzlibrary{positioning, decorations.pathreplacing} % Added
\usepackage{listings}

\usepackage{multirow}
\usepackage{soul}
\definecolor{hl}{rgb}{0.80,0.80,0.80} % Highlight color
\sethlcolor{hl}						  % Highlight
\usepackage{color}
\usepackage{color, colortbl}

\usepackage[nameinlink,capitalise]{cleveref} % Referencing

\newcommand{\eg}{\textit{e.g.,}}

\newcommand{\ie}{\textit{i.e.,}}

\newcommand{\wrt}{\textit{w.r.t.}}

\newcommand{\ka}{\texttt{K1}} % Added
\newcommand{\kb}{\texttt{K2}} % Added

\newcommand{\sumbox}[1]{\vspace{2mm}
\noindent\fbox{\parbox{0.98\linewidth}{\selectfont#1}}\\}

\tikzstyle{mybox} = [draw=black, thick, rectangle, inner ysep=5pt, inner xsep=5pt] % Summary box

\usepackage{pifont}

\AtBeginDocument{%
  \providecommand\BibTeX{{%
    \normalfont B\kern-0.5em{\scshape i\kern-0.25em b}\kern-0.8em\TeX}}}

\setcopyright{acmcopyright}
\copyrightyear{2024}
\acmYear{2024}
\acmDOI{XXXXXXX.XXXXXXX}

\acmConference[ICSE 2024]{46th International Conference on Software Engineering}{April 2024}{Lisbon, Portugal}
\acmPrice{15.00}
\acmISBN{978-1-4503-XXXX-X/18/06}

\begin{document}

%% The "title" command has an optional parameter, allowing the author to define a "short title" to be used in page headers.
\title{Evolutionary Generative Fuzzing  for Differential Testing  of the Kotlin Compiler}

%% The "author" command and its associated commands are used to define the authors and their affiliations. Of note is the shared affiliation of the first two authors, and the "authornote" and "authornotemark" commands used to denote shared contribution to the research.
\author{C\u alin Georgescu}
\orcid{0000-0002-8102-6389}
\affiliation{%
  \institution{Delft University of Technology}
  \city{Delft}
  \country{The Netherlands}
}
\email{C.A.Georgescu@tudelft.nl}

\author{Mitchell Olsthoorn}
\orcid{0000-0003-0551-6690}
\affiliation{%
  \institution{Delft University of Technology}
  \city{Delft}
  \country{The Netherlands}
}
\email{M.J.G.Olsthoorn@tudelft.nl}

\author{Pouria Derakhshanfar}
\orcid{0000-0003-3549-9019}
\affiliation{%
  \institution{JetBrains Research}
  \city{Amsterdam}
  \country{The Netherlands}
}
\email{Pouria.Derakhshanfar@jetbrains.com}

\author{Marat Akhin}
%\orcid{}
\affiliation{%
  \institution{JetBrains Research}
  \city{Amsterdam}
  \country{The Netherlands}
}
\email{Marat.Akhin@jetbrains.com}

\author{Annibale Panichella}
\orcid{0000-0002-7395-3588}
\affiliation{%
  \institution{Delft University of Technology}
  \city{Delft}
  \country{The Netherlands}
}
\email{A.Panichella@tudelft.nl}

%%
%% By default, the full list of authors will be used in the page
%% headers. Often, this list is too long, and will overlap
%% other information printed in the page headers. This command allows
%% the author to define a more concise list
%% of authors' names for this purpose.
%\renewcommand{\shortauthors}{Trovato and Tobin, et al.}

% Abstract
\begin{abstract}
  Compiler correctness is a cornerstone of reliable software development. However, systematic testing of compilers is infeasible, given the vast space of possible programs and the complexity of modern programming languages.
In this context, differential testing offers a practical methodology as it addresses the oracle problem by comparing the output of alternative compilers given the same set of programs as input.
In this paper, we investigate the effectiveness of differential testing in finding bugs within the Kotlin compilers developed at JetBrains. We propose a black-box generative approach that creates input programs for the \ka{} and \kb{} compilers.
First, we build workable models of Kotlin
semantic (semantic interface) and syntactic (enriched context-free grammar) language features, which are subsequently exploited to generate random code snippets. 
Second, we extend random sampling by introducing two genetic algorithms (GAs) that aim to generate more diverse input programs.
Our case study shows that the proposed approach effectively detects bugs in \ka{} and \kb{}; these bugs have been confirmed and (some) fixed by JetBrains developers. While we do not observe a significant difference w.r.t. the number of defects uncovered by the different search algorithms, random search and GAs are complementary as they find different categories of bugs. 
Finally, we provide insights into the relationships between the size, complexity, and fault detection capability of the generated input programs.

\end{abstract}

% Keywords



\maketitle

% Content
%%%%%%%%%%%%%%%%%%%%%%%%%%%%%%%%%%%%%%%%%%%%%%%%%%
\section{Introduction}
\label{sec:introduction}
%%%%%%%%%%%%%%%%%%%%%%%%%%%%%%%%%%%%%%%%%%%%%%%%%%

% Popularity, new compiler, relevance
% JetBrains expertise, creators of the language
% Kotlin Foundation numbers, 0.5M developers, 1M_ repositories
% Main developing language for Android

Compilers are an essential part of the software development ecosystem.
They allow developers to write programs in a high-level language, that can be understood by a human, and convert these to a format that machines can understand.
Kotlin is a popular upcoming high-level programming language that was developed by JetBrains in 2011, as an alternative to Java.
Currently, Kotlin is used by \num{15.9}M users and it is the main development language for Android.

Up until recently, Kotlin used the compiler that was introduced with the language when it was released, called \ka{}.
This compiler, however, is limited by its technical debt (\ie{} a consequence of software that expedited features over maintainability), making it harder to extend in the future.
Therefore, with the release of Kotlin~2.0, JetBrains is introducing an improved new compiler, called \kb{}.
As K2 is not just a refactored version of K1, but a complete rewrite of its frontend based on a different architecture, it is important to make sure that the two versions behave similarly.

Software verification~\cite{berard2013systems} is a specialized field of research that mathematically checks if a program complies with its requirements. However, it faces two main limitations when applied to compilers~\cite{chen2020survey}: (1) they do not scale to the size and complexity of compilers,
and (2) they require a model~(\ie{}~oracle) to determine if the output for a given input is correct.
%However, verification does not easily scale to the size of compilers and struggles with the test oracle problem.
%Checking a compiler for correct behavior requires a model~(\ie{}~oracle) to determine if the output produced by the compiler for a given input is correct.
%Creating such a model for a compiler is highly impractical and time-consuming.
An alternative approach that circumvents the oracle problem~\cite{barr2014oracle} is differential testing~\cite{mckeeman1998differential, chen2016coverage}.
Differential testing takes two different versions of the same program, supplies these with identical inputs, and compares their outputs. Eventual discrepancies highlight bugs in one of the two versions. 
%This allows for testing complex programs like compilers without having an exact model of their behavior.

%Differential testing can be classified in \textit{mutation-based} and \textit{generative}. The former relies on existing input programs for the compiler under test, which are 

Prior studies have successfully applied differential testing to find bugs in compilers for various programming languages, such as  Java~\cite{chen2019deep}, C/C++~\cite{yang2011finding,livinskii2020random}, and JavaScript~\cite{han2019codealchemist, holler2012fuzzing}. However, existing approaches either require an initial set of programs (seeds) to mutate~\cite{stepanov2021type} or generate programs relying on a context-free grammar (CFG) specification (\eg{} ~\cite{han2019codealchemist, holler2012fuzzing}). 
The latter approaches do not account for the rich semantic nuances often accompanying the grammar specifications. In the case of Kotlin, the CFG does not satisfy simple semantic rules without providing additional \textit{context}. This increases the likelihood of generating invalid code if no additional semantic constraints are considered.

This paper presents a case study of differential testing for the Kotlin compilers (\ka{} and \kb{} in particular) developed at JetBrains. We introduce a three-stage generative approach that generates valid Kotlin code, addressing the challenges accompanying the above-mentioned Kotlin specifications.
First, our approach truncates the Kotlin specifications and replaces them with two context-aware models, namely the \textit{enriched CFG} and the \textit{semantic interface}. The former is an enriched version of the CFG augmented with additional constraints. The latter encodes the meaning and the semantic relations between different code segments. 

Second, our approach implements two categories of search algorithms that generate valid Kotlin programs based on the context-aware models, namely random search (RS) and genetic algorithms (GAs). RS simply samples new Kotlin programs by randomly traversing the context-aware models. GAs, instead, evolve a pool (called \textit{population}) of randomly generated yet valid Kotlin programs with the aim of promoting/generating a diverse set of programs over time. We investigate two flavors of GAs, namely (1) a single-objective variant that maximizes the program diversity and (2) a many-objective variant that considers program size as well.

Our experimental results show that RS effectively detects differential bugs in the \ka{} and \kb{} compilers, namely \textit{out-of-memory errors} and \textit{resolution ambiguity}. Furthermore, GAs successfully detect further bugs related to \textit{conflicting overloads}. While RS and GA are statistically equivalent \wrt{} the number of compiler bugs they identify, they are complementary as they uncover different categories of bugs. 
We have reported three categories of bugs found by our approach, and they have been verified and confirmed by JetBrains developers. Some of these bugs have already been resolved in more recent compiler releases, while others are planned to be resolved in future releases.

This paper makes the following contributions:
\begin{itemize}
\item A three-stage generative approach that intertwines CFG, programming language semantics, and meta-heuristic search for differential testing of compilers.
\item A case study on the effectiveness of random search and meta-heuristics in testing Kotlin's \ka{} and \kb{} compilers.
\item The discovery and analysis of new differential bugs reported to and confirmed by JetBrains developers.
\item An in-depth analysis of the relationship between the characteristics of the generated programs (complexity and size) and their ability to uncover compiler bugs.
\item A replication package with code~\cite{kotfuzz-impl} and data~\cite{kotfuzz-data}.
\end{itemize}

% End: Generalizable approach, not Kotlin-specific

While our work focuses on Kotlin compilers, our approach can be applied to other compilers. Our study provides insights into the behavior of alternative search methods, which are valuable for compiler developers seeking to enhance the robustness and reliability of compilers across diverse programming languages.

% 1.5 Page max
%%%%%%%%%%%%%%%%%%%%%%%%%%%%%%%%%%%%%%%%%%%%%%%%%%
\section{Background and Related Work}
\label{sec:background}
%%%%%%%%%%%%%%%%%%%%%%%%%%%%%%%%%%%%%%%%%%%%%%%%%%

This section provides background information about Kotlin compilers and summarizes the related work in compiler testing and search-based software testing.

\subsection{Kotlin} % 0.5 col

Kotlin\footnote{\url{https://kotlinlang.org/}} is a relatively new language that was developed by JetBrains in 2011.
It is a general-purpose, high-level programming language that is statically typed and cross-platform.
JetBrains developed Kotlin as an alternative to Java, but now it runs not only on the JVM, but also on JavaScript, Native, and even WebAssembly.
Over the years, Kotlin has been gaining traction with the developer community.
According to the latest annual report\footnote{\url{https://www.jetbrains.com/lp/annualreport-2023/}}, Kotlin has \num{15.9}M users and \num{90} of the global top \num{100} companies use Kotlin.
In 2019, Google announced that Kotlin is the preferred language for Android.

The original compiler introduced with the language is called \ka{}.
This compiler has been updated with new features throughout the years, while at the same time accumulating technical and architectural debt.
Recently, JetBrains has been working on a new frontend for the compiler, called \kb{}\footnote{\url{https://blog.jetbrains.com/kotlin/2023/02/k2-kotlin-2-0/}}.
\kb{} aims to address the existing debt, speed up the development of new language features, improve the performance of the compiler, and fix bugs and inconsistencies in the compiler behavior.
Since the \kb{} compiler is not an iteration of the old compiler but a complete rewrite of its frontend component, it is important to ensure the two compilers behave similarly.

\subsection{Compiler Testing} % 0.75 col

% Compilers are ubiquitous and pivotal components
% at the core of innumerable software and programming language ecosystems.
% They translate the high-level, human-readable source code
% of computer programs into target-specific instructions that machines
% can parse and execute.
% The ramifications that arise from flawed
% compiler implementations can be both far-reaching and severe.
% The outstanding complexity and feature richness of modern compilers has created
% several formidable challenges, the scope of which is perhaps best exemplified
% by \citet{hoare2003verifying}
% declaring the task of building a provably correct compiler to be a 
% \textit{Grand Challenge} in computing research.

In recent decades, researchers and practitioners have invested tremendous resources to improve compiler testing through automation \cite{boujarwah1997compiler, chen2020survey}.
Modern approaches make use of Differential Testing~(DT) to heuristically assess different compilers of the same programming language.
They generate code snippets, which are used as the input for each compiler version with the goal of uncovering differences in behavior~\cite{mckeeman1998differential}.
Any difference in the compiler's outcome (\ie{} crashes, non-compiling snippets, or errors) highlights implementation errors that may affect real-world applications.
By making use of DT, we can circumvent the \textit{test oracle problem}~\cite{barr2014oracle}.

% In recent decades, researchers and practitioners have invested tremendous resources to improve compiler testing through automation.
% These approaches first generate code snippets to perform Differential Testing (DT) \cite{mckeeman1998differential}.
% DT is a technique that heuristically assesses multiple different compilers of the same programming language.
% The same piece of code is used as the input for each compiler version with the goal of uncovering differences in behavior.
% Differences in the outcome of compilers (\ie~crashes, miscompilations, or errors)
% highlight implementation errors that may affect real-world applications.

For this study, we make the distinction between \textit{test program generation} and \textit{program mutation} approaches as classified by \citet{chen2020survey}.
The former generates programs from scratch, without any external seed, while the latter mutates existing programs to generate new variations. 
Test program generation approaches also differ in their utilization of the language grammar.
\textit{Grammar-directed} approaches solely rely on the grammar to generate novel code snippets, whereas
\textit{grammar-aided} approaches heuristically exploit the grammar, which is often enriched with semantically rich context.

\citet{purdom1972sentence} devised one of the first algorithms aimed at testing the correctness of Context-Free Grammar (CFG) parsers.
They use a grammar-directed approach that applies iterative rewriting rules beginning with a starting symbol, to eventually exhaust the entire grammar specification.
\citet{yang2011finding} propose \textsc{Csmith}, a grammar-aided test program generation tool aimed at finding bugs in C compilers using DT.
\textsc{Csmith} fundamentally differs from Purdom's approach in that it follows a generation pattern that primarily focuses on the semantic properties of C rather than its grammar.

\citet{livinskii2020random} introduce \textsc{YARPGen}, a C/C++ program generation tool that aims to increase the expressiveness and diversity of test programs.
\textsc{YARPGen} proceeds in a top-down fashion and does not explicitly utilize a formal language grammar.
Instead, it tracks a \textit{type environment} that stores all visible composite data types, which implicitly guide the generative process.
The environment is iteratively enriched with newly generated types, each having access to all previous entries.
\citet{han2019codealchemist} introduce the notion of \textit{semantics-aware assembly}, which they implement in the \textsc{CodeAlchemist} tool, aimed at testing the JavaScript engine.
The key concept in semantics-aware assembly consists of building blocks, also referred to as \textit{code bricks}, a notion similar to \textsc{LangFuzz}'s \textit{code fragments}~\cite{holler2012fuzzing}.
A code brick consists of a valid JavaScript abstract syntax tree (AST) that is additionally annotated with an \textit{assembly constraint}.

% ISLa, *Smith below
\textit{Specification fuzzing} approaches envelop a third category of
code generation techniques that generally favor language-agnostic
declarative formulations over system-specific approaches.
The Input Specification Language (\textsc{ISLa}) \cite{steinhofel2022input}
is one such approach that allows users to annotate a grammar with semantic,
context-dependent constraints.
In this approach, users can introduce annotations that match the nuances
of real programming languages and thus circumvent implementing narrow-scoped fuzzers.
\textsc{ISLa} uses a \textit{solver} to iteratively expand the constrained
grammar specification, which allows it to generate code that is both semantically 
and syntactically aligned.
The Language Specification Language (\textsc{LaLa}) \cite{kreutzer2020language}
allows for declarative descriptions of attribute grammars \cite{knuth1968semantics}
as a means of introducing semantics to a context-free language.
The \textsc{LaLa} framework translates a given specification into Java classes
that are the input to the fuzzing process.
At runtime, the fuzzer instantiates ASTs that conform to the attribute
properties, while simultaneously checking for \textit{fail patterns}.
%\mitchell{Should this be failed patterns? A: No, that is term they use. I did not include the definition because it is quite involved and we do not have the space for it.}

Compared to the related work, we enhance the differential testing process by introducing heuristic-based guidance targeting the Kotlin language.
We use similar semantic modelling techniques as \textsc{Csmith}~\cite{yang2011finding} and \textsc{YARPGen}~\cite{livinskii2020random},
but we allow users of our tool to customize the root environment to a much greater extent.
Our implementation enables users to include arbitrary Kotlin code that is automatically parsed and integrated into the semantic context.
During sampling, we repeatedly query the context and use its constituents (\ie{} user-provided classes) to generate new code.
Similar to \textsc{CodeAlchemist} \cite{han2019codealchemist}
and \textsc{LangFuzz} \cite{holler2012fuzzing}, we structurally decompose code.
However, in contrast to those approaches, we only employ decomposition
as a means of adding variation to our code during generation
rather than extracting information from existing code bases.
We also decided against using \textsc{ISLa}
\cite{steinhofel2022input} and \textsc{LaLa} \cite{kreutzer2020language} because of
the significant effort required to encode the rich semantics of
Kotlin within their respective frameworks.
Additionally, the fuzzers of both approaches heavily rely on an enumeration of random expansions,
whereas our approach enables a more sophisticated heuristic search.
% Our approach structurally decomposes code in a manner similar to \textsc{CodeAlchemist} \cite{han2019codealchemist} and \textsc{LangFuzz} \cite{holler2012fuzzing}, however, we employ this decomposition as a means of adding variation to our code during generation rather than extract information from existing code bases.
Finally, \citet{stepanov2021type} also targets Kotlin with their mutation-driven method, which, unlike our approach, relies on existing code as a starting point for the fuzzer.

\subsection{Search-Based Software Testing} % 0.75 col
% 1 Page Max
Search-based Software Testing (SBST) is a broad umbrella of approaches and tools aimed at automating the process of generating test cases~\cite{mcminn2011search}. The automation is achieved by utilizing optimization algorithms (e.g., genetic algorithms) that iteratively evolve a set of randomly generated test cases toward optimizing given testing criteria (e.g., branch coverage)~\cite{anand2013orchestrated}.
To this aim, meta-heuristics leverage a fitness function (or objective) to evaluate how closely the test execution aligns with those goals.
Depending on which type of information the fitness function relies on, SBST techniques can be classified in \textit{white-box} and \textit{black-box}. The former techniques use the internal information (e.g., coverage data) of the program under test to assess the ``fitness'' of the generated tests~\cite{mcminn2011search}. 
The latter techniques do not require access to the source code (bytecode) but rely on external information~\cite{corradini2021empirical}, such as input diversity~\cite{aghababaeyan2023black}.

Previous research has demonstrated the effectiveness of SBST across various testing levels, including unit~\cite{fraser2011evosuite, panichella2015reformulating, panichella2017automated}, integration~\cite{derakhshanfar2022generating}, system levels~\cite{arcuri2019restful, corradini2021empirical}, and concurrency testing~\cite{van2023evolutionary}. SBST approaches have shown to be particularly promising, outperforming random testing, in achieving extensive coverage~\cite{panichella2021sbst, campos2018empirical}, detecting defects~\cite{fraser20151600, derakhshanfar2022generating}, or testing cyber-physical systems~\cite{abdessalem2018testing}.

In this work, we aim to investigate the effectiveness of genetic algorithm and random search when applied to the detection of bugs in Kotlin compilers using differential testing. As we illustrate in the next section, our approach is black-box, thus focusing on the input and output of the compilers under test.
%%%%%%%%%%%%%%%%%%%%%%%%%%%%%%%%%%%%%%%%%%%%%%%%%%
\section{Approach}
\label{sec:approach}
%%%%%%%%%%%%%%%%%%%%%%%%%%%%%%%%%%%%%%%%%%%%%%%%%%

Our approach aims to automatically generate 100\% valid code
that uncovers defects in the Kotlin compilers.
To achieve this goal, we designed a multi-stage approach comprising three phases, as depicted in \cref{fig:overview}.
Phase I consists of building workable models of Kotlin semantic and syntactic language features, which are subsequently exploited to generate (\textit{sample}) random code snippets.
We elaborate on this step in \Cref{subsec:phase1}.
\Cref{subsec:phase2} covers Phase II, in which we provide \textit{guidance} to the random sampling process by introducing search objectives and utilizing Evolutionary Algorithms (EAs).

Phase III performs \textit{differential testing} (DT) on \ka~and \kb~using the output of Phase II as compiler input.
In this final step, we differentiate between three different cases.
If both compilers display the same behavior (\ie{} they both successfully compile the code), no defects have been uncovered.
If the two compilers give different verdicts, (\ie{} one compiles the code while the other raises an error, as shown in \cref{fig:overview}), then one of the versions under test is erroneous.
Finally, if either of the compilers \textit{crashes}, we individually analyze the cause of the crash through the compiler's logging mechanism.

% Overview Diagram w/ steps
\tikzset{XOR/.style={draw,circle,append after command={
        [shorten >=\pgflinewidth, shorten <=\pgflinewidth,]
        (\tikzlastnode.north) edge (\tikzlastnode.south)
        (\tikzlastnode.east) edge (\tikzlastnode.west)
        }
    }
}
\tikzset{line/.style={draw, -latex',shorten <=1bp,shorten >=1bp}}

\def\layersep{0.5}
\def\nodesep{0.5}

\newcommand{\cmark}{\ding{51}}%
\newcommand{\xmark}{\ding{55}}%

\colorlet{gray1}{gray!40!black}
\colorlet{gray2}{gray!80!black}

\begin{figure*}[!hbtp]
\vspace{2cm}
% \resizebox{0.005\columnwidth}{!}{
\begin{tikzpicture}[transform canvas={scale=0.9},node/.style={circle, draw, thick}]

\foreach \Z in {0,0.5,1,1.5}
 {\draw[fill=gray,draw=black] (-8,0,\Z) rectangle (-7.25,1,\Z);}

\node[] at (-7.5,1.5)   (a) {Kotlin Files};

 \draw [-stealth] (-7.15, 0.5) -- node[pos=0.5,above]{Extraction} (-5.5, 0.5);
 
\pgfmathsetmacro{\cubex}{1.25}
\pgfmathsetmacro{\cubey}{1.25}
\pgfmathsetmacro{\cubez}{1.25}
\draw[black,fill=gray] (-4,1,0) -- ++(-\cubex,0,0) -- ++(0,-\cubey,0) -- ++(\cubex,0,0) -- cycle;
\draw[black,fill=gray] (-4,1,0) -- ++(0,0,-\cubez) -- ++(0,-\cubey,0) -- ++(0,0,\cubez) --  cycle;
\draw[black,fill=gray] (-4,1,0) -- ++(-\cubex,0,0) -- ++(0,0,-\cubez) -- ++(\cubex,0,0) -- cycle;

\node[] at (-4.25,2.175)   (a) {\small (Semantics)};
\node[] at (-4.25,1.875)   (a) {Context};

% \draw[] (-1.5, -0.5) rectangle (1.25, 1.5);

 %%%%%%%%%%%%%%%%%%%%%%%%%%%% Bottom half
 \foreach \Z in {0,0.5,1,1.5}
 {\draw[fill=gray,draw=black] (-8,-2,\Z) rectangle (-7.25,-1,\Z);}
\node[] at (-7.5,-3)   (a) {Kotlin CFG};

\draw [-stealth] (-7.15, -2) -- (-5.5, -2) node[pos=0.5,above]{Enrichment};

  \foreach \y in {1,...,3}{
      \node[node] (i\y) at (-5,\nodesep*\y-3) {};
      \node[node, right=\layersep of i\y] (h1\y) {};
      \node[node, right=\layersep of h1\y] (h2\y) {};
    }

  \foreach \source in {1,...,3}
  \foreach \dest in {1,...,3}{
      \path[-stealth, thick] (i\source) edge (h1\dest);
      \path[-stealth, thick] (h1\source) edge (h2\dest);
    }

\draw[] (-5.25, -2.75) rectangle (-3, -1.25);

\node[] at (-4.25,-3)   (a) {Enriched CFG};
\node[] at (-4.25,-3.3)   (a) {\small (Syntax)};
%%%%%%%%%%%%%%%%%%%%%%%%%%%%%%%%% Middle section
\node (XOR-aa)[XOR,scale=2] at (-2.5,-0.75) {};
\draw [-stealth] (-2.75, -2) -- (-2.5, -2) -- (-2.5,-1.05);
\draw [-stealth] (-3.25, 0.5) -- (-2.5, 0.5) -- (-2.5,-0.45);
\draw [-stealth] (-2.15, -0.75) -- (-1.9, -0.75);

\node[] at (-1.25,-0.75)   (rs) {RS};
\draw[] (-1.75, -1.25) rectangle (-0.75, -0.25);
\draw [-stealth] (-0.6, -0.75) -- (-0.15, -0.75);

\foreach \Z in {0,0.5,1,1.5}
 {\draw[fill=gray,draw=black] (0.5,-1,\Z) rectangle (1.25,0,\Z);}
\node[] at (0.65,0.5)   (a) {Kotlin Code};
\draw [-stealth] (1.5, -0.75) -- (1.85, -0.75);

\node[] at (2.5,-0.75)   (ea) {EA};
\draw[] (2, -1.25) rectangle (3, -0.25);
\draw [-stealth] (2.5, -1.35) -- (2.5, -2) -- (0.65,-2) -- (0.65, -1.65);
\node[] at (1.575,-1.675)   (a) {Variation};
\draw [-stealth] (3.15, -0.75) -- (3.75, -0.75);

\foreach \Z in {0,0.5,1,1.5}
 {\draw[fill=gray,draw=black] (4.5,-1,\Z) rectangle (5.25,0,\Z);}
\node[] at (4.75,0.5)   (a) {Kotlin Code};
\draw [-stealth] (5.5,-0.75) -- (5.75,-0.75) -- (5.75,0.5) -- (6,0.5);
\draw [-stealth] (5.5,-0.75) -- (5.75,-0.75) -- (5.75,-2) -- (6,-2);

\pgfmathsetmacro{\cubexc}{0.75}
\pgfmathsetmacro{\cubeyc}{0.75}
\pgfmathsetmacro{\cubezc}{0.75}
%%% K1
\draw[black,fill=gray] (7,0.75,0) -- ++(-\cubexc,0,0) -- ++(0,-\cubeyc,0) -- ++(\cubexc,0,0) -- cycle;
\draw[black,fill=gray] (7,0.75,0) -- ++(0,0,-\cubezc) -- ++(0,-\cubeyc,0) -- ++(0,0,\cubezc) --  cycle;
\draw[black,fill=gray] (7,0.75,0) -- ++(-\cubexc,0,0) -- ++(0,0,-\cubezc) -- ++(\cubexc,0,0) -- cycle;
\node[] at (6.85,1.375)   (a) {\texttt{K1}};
\draw[-stealth] (7.5,0.5) -- node[pos=0.5,above]{Compile} (8.5,0.5);
\node[] at (8.75,0.5)   (vcheck) {\Huge\cmark};

%%% K2
\draw[black,fill=gray] (7,-1.75,0) -- ++(-\cubexc,0,0) -- ++(0,-\cubeyc,0) -- ++(\cubexc,0,0) -- cycle;
\draw[black,fill=gray] (7,-1.75,0) -- ++(0,0,-\cubezc) -- ++(0,-\cubeyc,0) -- ++(0,0,\cubezc) --  cycle;
\draw[black,fill=gray] (7,-1.75,0) -- ++(-\cubexc,0,0) -- ++(0,0,-\cubezc) -- ++(\cubexc,0,0) -- cycle;
\node[] at (6.85,-2.915)   (a) {\texttt{K2}};
\draw[-stealth] (7.5,-2) -- node[pos=0.5,above]{Compile} (8.5,-2);
\node[] at (8.75,-2)   (vcheck) {\Huge\xmark};

% \node[] at (6.85,1.375)   (a) {\texttt{K1}};
% Alternative: draw loop inside of EA box
% \node[] at (3,0.75)   (pop) {Population};

% \node[] at (3,-0.75)   (par) {Parents};

% \node[] at (3,-2.25)   (par) {Offspring};

%%%%%%%%%%%%%%%%%%%%%%%%%%%%%%% Phases
\draw[densely dotted] (-0.375, 2) -- (-0.375,-3.5);
\draw[densely dotted] (3.45, 2) -- (3.45,-3.5);
\draw [decorate,
    decoration = {brace,mirror}] (-8.5,-3.5) --  (-0.425,-3.5);
\draw [decorate,
    decoration = {brace,mirror}] (-0.325,-3.5) --  (3.4,-3.5);
\draw [decorate,
    decoration = {brace,mirror}] (3.5,-3.5) --  (9,-3.5);
    
\node[] at (-4.4625,-3.75) (p1) {Phase I - Modeling and Sampling};
\node[] at (1.5375,-3.75) (p2) {Phase II - Search};
\node[] at (6.25,-3.75) (p3) {Phase III - Differential Testing};

\end{tikzpicture}
% }
\vspace{3.3cm}
\caption{Overview of our approach.}
\label{fig:overview}
\end{figure*}
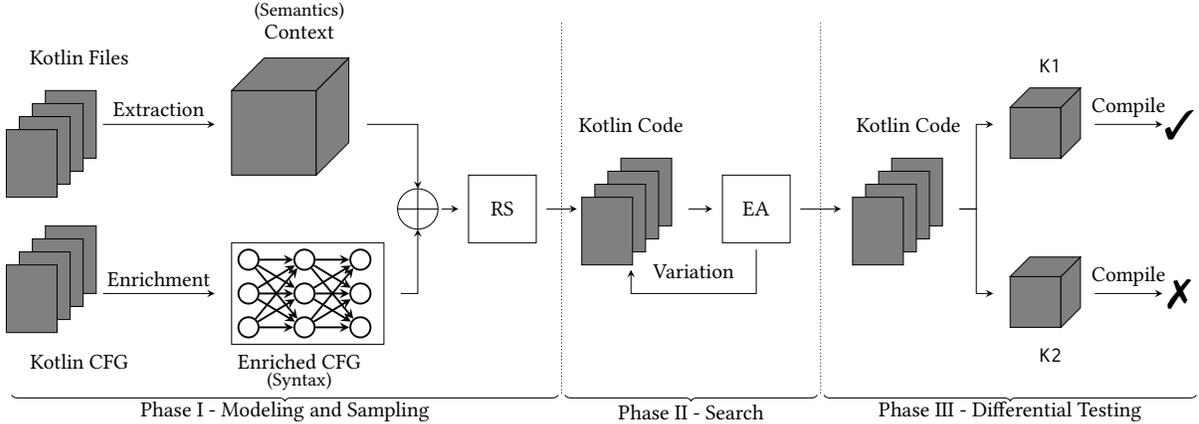

\subsection{Random Code Generation \label{subsec:phase1}}

In this subsection, we delve into the models that our approach uses to encompass the structure (syntax) and the meaning (semantics) of programs.
These models serve as the basis for generating valid Kotlin code and form the foundation of the EAs in Phase II.

\subsubsection{Syntactic Interface}
Kotlin is a rich programming language, the structure of which is defined in terms of a formal Context-Free Grammar (CFG)~\cite{kotlingrammar}.
The language developers made a full grammar specification available in the popular \textsc{antlr4} \cite{parr2013definitive} parser-generator framework.
Though crucial for verifying the structural soundness of Kotlin code, the CFG poses several practical challenges.

With over 480 symbols and 700+ productions, the Kotlin CFG is a complex aggregation of structural relations.
However, the inherent lack of context of the grammar gives rise to several obstacles that make straightforward generation algorithms impractical.
For instance, classic grammar-based fuzzers such as the one introduced by Purdom \cite{purdom1972sentence} do not account for the rich semantic nuances often accompanying CFG specifications.
In the case of Kotlin, the CFG cannot satisfy even simple semantic rules such as \textit{a variable name must be defined before assignment}.
When paired with the significant complexity of the specification, this causes any attempt to sample the grammar independently to result in invalid code with an overwhelming likelihood, due to the accumulation of semantic constraints that are unaccounted for in the grammar.

To address this limitation, we implemented two key operations that modify the standard Kotlin CFG.
First, we selectively \textit{truncate} the grammar at points in the specification where significant semantic nuance occurs.
This operation preserves the \textit{shape} in the grammar (\ie{} no new symbols or rules are created, but transitions between symbols are retained), while drastically reducing its complexity and increasing the likelihood of generating valid code.
Second, we endow each grammar rule with a \texttt{sample} property that overrides the standard specification of the symbol in the CFG.

We only perform this latter operation on semantically rich symbols that remain in the CFG's composition after the truncation step.
Simpler nodes (such as \textit{cardinality} or \textit{optionality}) do not require specific \texttt{sample} implementations and instead fall back on their default specification.
The fallback mechanism guarantees that grammar-abiding options are always available, even if individual \texttt{sample} rules are not implemented.
This, in turn, enables a trade-off between specificity and implementation complexity.

The purpose of the \texttt{sample} property is to ensure that grammar traversal algorithms additionally account for constraints not included in the grammar specification.
Together, the two operations simplify and augment the Kotlin CFG in a process we call \textit{enrichment} (\Cref{fig:overview}, Phase I).
The output of this transformation is an annotated Directed Acyclic Graph (DAG), where nodes correspond to symbols and edges pertain to rules in the original grammar.

\subsubsection{Semantic Interface}

To effectively utilize the syntactic interface of the enriched CFG, our method requires a corresponding semantic counterpart that encodes the meaning and semantic relation between different code segments.
The abstraction that encompasses these concepts is hereafter referred to as \textit{semantic context}.

Our implementation of the semantic context fulfills three key requirements within the fuzzer.
Firstly, it furnishes the enriched CFG with a mechanism for querying useful information from the available code.
This  allows traversal algorithms to discern which productions are feasible.
Secondly, the context actively tracks changes to the semantics of a program as it is generated.
This \textit{mutability} enables an iterative increase in the complexity of the sampled code, as previously generated code (\ie~variables) can appear again in later lines.
Finally, we construct the semantic context through an \textit{extraction} process, that parses and extracts information from provided Kotlin files.
This process ensures the versatility of our approach, as users can provide arbitrary files as input to constrain (or broaden) the generative process.

The context data structure tracks three traits of Kotlin programs.
First, it maintains a set of visible \textit{callables} within reachable code.
We use the term \textit{callables} in a similar fashion to \cite{stepanov2021type}, encompassing all visible functions, properties, constructors, variables, constants, and primitives provided to the fuzzer.
This data is stored in a $\lambda$-calculus-like representation that
captures their properties.
The context additionally accounts for the \textit{type hierarchy} of its programs and the constraints related to parameterized types.
Lastly, various semantic constraints are embedded within the queries that the context supports, including both universal Kotlin rules and context-sensitive restrictions
(\ie~\textit{only sample types that at least one callable in the context can return}).

\subsubsection{Random Sampling/Search}

\begin{algorithm}[t]
\small
	\SetKwData{Left}{left}
	\SetKwData{This}{this}
	\SetKwData{Up}{up}
	\SetKwFunction{init}{InitializeAndEvaluatePopulation}
	\SetKwFunction{terminate}{ShouldTermiante}
	\SetKwFunction{offspring}{CreateAndEvaluateOffspring}
	\SetKwFunction{select}{SelectIndividuals}
	\SetKwFunction{time}{TimeElapsed}
	\SetKwFunction{clone}{Clone}
	\SetKwFunction{sample}{sample}
	\SetKwInOut{Input}{Input}
	\SetKwInOut{Output}{Output}
	\Input{CFG $\mathcal{N}$, Context $\mathcal{C}$, Runtime $s$}
	% \Output{Collection of generated Kotlin files}
	\BlankLine
	\DontPrintSemicolon
	%\emph{special treatment of the first line}\;
	$A \gets \emptyset$\;
	\While{$\neg$ \time{$s$}}{
		$c \gets \clone {$\mathcal{C}$}$\;
		$b \gets \sample {$\mathcal{N}$, c}$\;
		$A \gets A \cup \{ b \}$\;
	}
	\Return $A$\;

	\caption{Random Sampling}
	\label{alg:rs}
\end{algorithm}
% RS and algorithm

The final step in Phase I combines the enriched CFG
and the semantic context in a straightforward manner that produces
random semantically valid Kotlin code.
To do so, we follow the enriched grammar structure,
randomly selecting transitions between nodes,
while simultaneously querying the attached semantic context.
We refer to this procedure interchangeably as Random Sampling or Random Search (RS).

\Cref{alg:rs} outlines RS.
The algorithm begins by initializing an archive $A$ (line 1)
that tracks all generated code.
The main loop (lines 2-5) proceeds by
first creating a clone of the root context (line 3) before
querying the \texttt{sample} property of the given target CFG node (line 4).
The purpose of the context clone is to ensure that changes performed
in a sampling round do not propagate to later, independent samples.
RS then adds the obtained sample to the archive (line 5) and 
returns the collected samples (line 6).
%Both the context and the node that form the root of the generation
%process serve as input to RS, allowing for flexible user customization.

\subsection{Evolutionary Fuzzing \label{subsec:phase2}}

Though effective at drawing random samples of valid Kotlin code,
\Cref{alg:rs} lacks direction.
Specifically, it lacks a mechanism that would allow sampled code
to undergo structural changes that increase the likelihood of
uncovering compiler defects.
To introduce such a mechanism, we rely on EAs as a framework
of performing iterative changes on randomly
sampled Kotlin code in Phase II of our approach.
We first address the genetic representation of code,
before describing the variation operators
and finally, formulating the overarching EA in detail.

\subsubsection{Solution Encoding}
The individuals that make up the population of the EA
consist of independently valid pieces of code,
hereafter referred to as \textit{blocks}.
The goal of this denomination is to isolate pieces of code
that are both syntactically and semantically \textit{self-contained}
(\ie~that have no external dependencies).
Each block $B$ is comprised of an ordered set of \textit{snippets}
$[s_1, ..., s_n]$, representing isolated
pieces of Kotlin code with inner scopes (\ie~classes, functions).
We order snippets based on the topology of their dependencies
to more efficiently detect and extract self-contained structures during search. 
Each snippet $s$ is itself a 4-tuple $\langle N, \Lambda, D, F \rangle$
with $N$ the name of the snippet,
$\Lambda$ the $\lambda$-calculus formatted metadata which encodes the snippet's input-output behavior,
$D$ a list of snippets $s$ depends on,
and $F$ an ordered list of \textit{fragments}, the lowest denomination of our representation.
Fragments are simply text-encoded pieces of Kotlin code that generally correspond
to single lines of code.
These generally include statements and expressions.

This hierarchical representation provides two key advantages.
First, it captures the structural composition of the underlying Kotlin program
with increasing levels of granularity.
This enables the design of variation operators that
perform complex alterations to the structure of sampled programs,
which sampling alone cannot.
Second, it allows for automated dependency and conflict analysis.
Since snippets include metadata regarding their signatures and dependencies,
it is straightforward to reason about the conflicts that may arise when including
two snippets in the same block or about the requirements
that removing a snippet may entail.
Fragments include no such metadata, reducing overhead.
The dependencies of a snippet are equivalent to the cumulative
dependencies of all of the snippet's fragments.
In our formulation, fragments are bound to a single snippet's scope
and are not subject to any further variation during the search.
Because of this, reasoning about the snippet's dependencies
suffices to ensure the validity of the code.

\subsubsection{Variation Operators}
\label{sec:operators}
Before describing the variation operators of our EA,
we first establish the notion of a \textit{self-sufficient partition} of a block.
Given a block $B = [s_1, ..., s_n]$, we can obtain a partition $B' = [s_1', ..., s_m']$
starting from a snippet $s_i = s_1' \in B$ by including (i) $s_i$,
(ii) all snippets that $s_i$ directly or indirectly depends on,
(iii) all snippets that directly or indirectly depend on $s_i$,
and (iv) recursively performing selections (ii) and (iii).
The new  partition $B'$ is a self-sufficient block,
as it has no external dependencies.

Using this notion, we define three mutation operators that perform changes on arbitrary blocks $B$.
The \textit{removal} operator first selects a random snippet $s_r \in B$ and removes its
corresponding self-sufficient partition $B_{s_r} \subseteq B$.
The \textit{\underline{c}ontext-\underline{f}ree \underline{a}ddition operator} samples a new block $B_{\mathit{cfa}}$
from the root context and appends its snippets to $B$.
Finally, the \textit{\underline{c}ontext-\underline{a}ware \underline{a}ddition operator} first merges the context
of $B$ with the root context before performing a sample operation that results in a new block $B_{\mathit{caa}}$.
The mutation consists of appending the snippets of $B_{\mathit{caa}}$ to $B$.

We additionally implement a recombination operator that takes as input two parent blocks
$B_{p_1}$ and $B_{p_2}$ and swaps two self-sufficient partitions $B_{x_1} \subseteq B_{p_1}$
and $B_{x_2} \subseteq B_{p_2}$ to obtain two new offspring blocks
$B_{o_1} = (B_{p_1} - B_{x_1}) \cup B_{x_2}$ and
$B_{o_2} = (B_{p_2} - B_{x_2}) \cup B_{x_1}$.
We perform conflict analysis prior to recombination
such that we only select pairs of blocks with no conflicting signatures.

\subsubsection{Heuristics} % Algorithms for RS, SODGA, MODGA

We bring together the tools developed in this section in two formulations of
genetic algorithms (GAs).
Both algorithms attempt to optimize the \textit{diversity} of sampled code under
the hypothesis that structurally diverse code is more likely
to stress different components of the compiler and thus
uncover more defects.

Each algorithm implements a different measure of diversity through the fitness
function involved in the selection mechanism.
To map individuals to a numerical fitness space, we first establish
a notion of \textit{similarity} between blocks.
For any two blocks $B_1$ and $B_2$ we define a mapping $m : \mathcal{B} \to \mathbb{N}^k$
that transforms blocks of Kotlin code from the abstract space $\mathcal{B}$ to
$k$-dimensional natural number vectors.
Each position of the vector represents the number of times a particular
Kotlin language feature (\ie~\texttt{if} expressions, functions)
appears in the input block, capturing a crude estimation of the structural
composition of the underlying program.
Using this mapping, any common measure of distance 
$d:\mathbb{N}^k \times \mathbb{N}^k \to \mathbb{R}$ (\ie{} euclidean norm) can
be used to determine the similarity between two blocks.
Using these notions, we define the \textit{population-wide dissimilarity}
of a block in $B$ in population $P$ according to \Cref{eq:dis}:
\begin{equation}
dis(B, P) = \min_{B_i \in P - \{ B \}} \left\{ d\left(m(B), m(B_i)\right) \right\}
\label{eq:dis}
\end{equation}

Intuitively, \Cref{eq:dis} measures the distance between $B$
and its most similar individual in $P$.
We use this formula to construct the population-wide diversity fitness function
in \Cref{eq:sof}.
We define $f^{(SO)}_{\texttt{DIV}}$ as a single-objective (SO)
fitness function, which seeks to minimize a value proportional to the inverse
of \Cref{eq:dis}, effectively maximizing dissimilarity:

\begin{equation}
\small
\min_{B \in \mathcal{B}} f^{(SO)}_{\texttt{DIV}}(B, P) = \frac{1}{1 + dis(B, P)}
\label{eq:sof}
\end{equation}

\begin{algorithm}[t]
\small
	\SetKwData{Left}{left}
	\SetKwData{This}{this}
	\SetKwData{Up}{up}
	\SetKwFunction{init}{InitializeAndEvaluatePopulation}
	\SetKwFunction{terminate}{ShouldTermiante}
	\SetKwFunction{offspring}{CreateAndEvaluateOffspring}
	\SetKwFunction{select}{SelectIndividuals}
	\SetKwFunction{time}{TimeElapsed}
	\SetKwFunction{clone}{Clone}
	\SetKwFunction{sample}{Sample}
	\SetKwInOut{Input}{Input}
	\SetKwInOut{Output}{Output}
	\Input{Population size $n$, CFG $\mathcal{N}$, Context $\mathcal{C}$, Runtime $s$}
	% \Output{Collection of generated Kotlin files}
	\BlankLine
	\DontPrintSemicolon
	$t \leftarrow 1$\;
	$P_1, P^{*} \leftarrow $ \init{$n, \mathcal{N}, \mathcal{C}$}\;
	\While{$\neg$ \time{$s$}}{
		$O_t \leftarrow$ \offspring {$P_t$}\;
		$P_{t+1} \leftarrow$ \select{$P_{t}, O_{t}, f^{(SO)}_{\texttt{DIV}}$}\;
		\If{$\sum_{b\in P_{t+1}} f^{(SO)}_{\texttt{DIV}}(b, P_{t+1}) > \sum_{b\in P^{*}} f^{(SO)}_{\texttt{DIV}}(b, P^{*})$} {
		$P^{*} \leftarrow P_{t+1}$\;
		}
		$t \leftarrow t + 1$\;
	}
	\Return $P^{*}$

	\caption{Single-Objective Diversity-based GA}
	\label{alg:sodga}
\end{algorithm}

\Cref{alg:sodga} describes the Single Objective Diversity Genetic Algorithm (SODGA)
that optimizes $f^{(SO)}_{\texttt{DIV}}$.
The algorithm begins by initializing a counter
and instantiating the population by means of RS (lines 1, 2).
In addition to the standard population, SODGA also tracks the most diverse population
encountered during the run in the variable $P^*$.
Next, the algorithm proceeds in a loop (lines 3-8) where
the variation operators give rise to offspring (line 4)
before triaging the population through selection (line 5).
In each generation, the most diverse population is updated if
a better (more diverse) set of individuals emerges (lines 6, 7).
After the time budget has been exhausted, the algorithm returns
the population $P^*$.

The fitness function of SODGA is fundamentally
dependent on each generation's population, which prevents
it from ever converging to a stable solution set.
While this is desirable for lengthy fuzzing campaigns, we propose a
second algorithm that aims to provide a stable converging behavior.
In contrast to SODGA, we design this alternative as a many-objective (MO) approach
that seeks to construct a stable archive of diverse yet small Kotlin programs.
\Cref{eq:mof} describes the MO fitness function that attempts to simultaneously
minimize the size of generated programs ($f_{sz}$) and maximize the number
of each language features present in them ($f_{l_i}$ from an abstract
set of language features $\mathbb{L}$):
\begin{equation}
\small
\max_{B \in \mathcal{B}} f^{(MO)}_{\texttt{DIV}}(B) = \left\lbrace -f_{sz}(B), ~f_{l_1}(B),~\dots,~f_{l_n}(B)|~l_i \in \mathbb{L} \right\rbrace
\label{eq:mof}
\end{equation}
In contrast to the SO approach, the size objective in MODGA favors small programs, which helps isolate uncovered defects.

\Cref{alg:modga} describes the implementation of the MO Diversity GA (MODGA).
It first initializes an elitist archive (line 1) before sampling a random initial
population (line 2).
The main loop proceeds in standard GA fashion, with an additional
archive update step (line 4) that processes newly generated files.
Selection (line 6) is carried out by means of Pareto-domination counting \cite{marler2004survey}.
MODGA returns the elements of the archive at the end of the run.

\begin{algorithm}[t]
\small
	\SetKwData{Left}{left}
	\SetKwData{This}{this}
	\SetKwData{Up}{up}
	\SetKwFunction{init}{InitializeAndEvaluatePopulation}
	\SetKwFunction{terminate}{ShouldTermiante}
	\SetKwFunction{offspring}{CreateAndEvaluateOffspring}
	\SetKwFunction{select}{DominationSelection}
	\SetKwFunction{time}{TimeElapsed}
	\SetKwFunction{processarchive}{ProcessNewArchiveEntries}
	\SetKwFunction{initarchive}{InitializeElitistArchive}
	\SetKwFunction{sample}{Sample}
	\SetKwInOut{Input}{Input}
	\SetKwFunction{shuffle}{Shuffle}
	\SetKwInOut{Output}{Output}
	\Input{Population size $n$, CFG $\mathcal{N}$, Context $\mathcal{C}$, Runtime $s$}
	% \Output{Collection of generated Kotlin files}
	\BlankLine
	\DontPrintSemicolon
	$A \leftarrow$ \initarchive{$f^{(MO)}_{\texttt{DIV}}$}; $t \leftarrow 1$\;
	
	$P_1 \leftarrow $ \init{$n, \mathcal{N}, \mathcal{C}$}\;
	\While {$\neg$ \time{$s_t$}}{
		$A \leftarrow$ \processarchive{$P_t, f^{(MO)}_{\texttt{DIV}}$}\;
		$O_t \leftarrow$ \offspring {$P_t$}\;
		$P_{t+1} \leftarrow$ \select{$P_{t}, O_{t}, f^{(MO)}_{\texttt{DIV}}$}; \;
		$t \leftarrow t + 1$\;
	}
	\Return $A$\;
	\caption{Many-Objective Diversity-based GA}
	\label{alg:modga}
\end{algorithm}
% 3 Pages = columns
%%%%%%%%%%%%%%%%%%%%%%%%%%%%%%%%%%%%%%%%%%%%%%%%%%
\section{Empirical Study}
\label{sec:study}
%%%%%%%%%%%%%%%%%%%%%%%%%%%%%%%%%%%%%%%%%%%%%%%%%%

Our empirical study aims to evaluate RS, SODGA, and MODGA \wrt{} 
their capability of uncovering bugs in the Kotlin compiler.
We begin by examining the impact of internal parameter 
values on each algorithm to gain insights into their behavior. 
Specifically, we  analyze how the notion of 
\textit{expression simplicity} influences block generation in RS. 
Additionally, we investigate the role of the chosen distance 
metric in \Cref{eq:dis} on the performance of SODGA.
Furthermore, we conduct a comparative analysis of the performance of different
algorithms in terms of the number
of defects they find during each run.
We summarize these goals within the following research questions:

\begin{questions}
    \item \textit{How does expression simplicity impact the properties of files generated by RS?}
    \item \textit{How does the distance measure influence the properties of
files generated by SODGA?}
    \item \textit{How effective are RS, SODGA, and MODGA in terms of uncovering bugs in the Kotlin
compiler?}
\end{questions}

\subsection{Framework}
We implemented the methods described in this paper in an open-source repository%
\footnote{https://github.com/ciselab/kotlin-compiler-fuzzer}
that includes additional heuristics not detailed in this study.
In addition to the fuzzer implementation, it contains an extensive framework for 
thorough customization and analysis of heuristics.
Furthermore, it provides utilities for replicating the results of this 
study \cite{kotfuzz-impl, kotfuzz-data}.
This includes functionality that automates DT and defect classification, 
as well as the aggregation of data related to the fuzzer's performance.

\begin{table}[t]
    \centering
    \caption{Overview of evaluated algorithms.}
    \label{tab:config}
    \small
    \begin{tabular}{ccccc}
    \toprule
    Name & Fitness & Objectives & Selection & Parameter \\
    \midrule
    RS & - & - & - & Simplicity Bias ($p_b$) \\
    \midrule
    SODGA & $f^{(SO)}_{\texttt{DIV}}$ & 1 & Tournament & Distance ($l^2$ or $l^\infty$)\\
    \midrule
    MODGA & $f^{(MO)}_{\texttt{DIV}}$ & $\mid \mathbb{L} \mid + 1 = 7$ & Dom. Rank & -\\
    \bottomrule
    \end{tabular}
\end{table}

\subsection{Configurations}

We assess the implementation of three distinct algorithms: RS, SODGA, and MODGA.
All implementations operate on a subset of the Kotlin CFG
that includes five language features: function declarations\footnote{\url{https://kotlinlang.org/spec/declarations.html\#declarations}}, statements\footnote{\url{https://kotlinlang.org/spec/statements.html\#statements}},
assignments\footnote{\url{https://kotlinlang.org/spec/statements.html\#assignments}}, 
and four types of expressions\footnote{\url{https://kotlinlang.org/spec/expressions.html\#expressions}}.
This means the output of \Cref{eq:mof} is a seven-dimensional vector.
The first entry of the vector denotes the program size, and the remaining six entries
count the frequency (number) of each language feature in the corresponding block.
%Though limited, this subset of the Kotlin language suffices to highlight the
%individual traits of each heuristic.

\Cref{tab:config} lays out the overview of the configuration
for each evaluated algorithm.
To answer RQ1, we introduce the \textit{simplicity bias} parameter
that governs the complexity of sampled blocks.
This parameter operates within the \texttt{sample}
property of CFG nodes for the expressions.
The bias influences the probability
of sampling \textit{simple} expressions (\ie~function calls and property accesses)
as opposed to more complex counterparts (\ie~\texttt{if}-expressions).
The simplicity bias is expressed as a number $p_b \in [0, 1]$,
where $p_b$ indicates the probability of sampling \textit{simple} expressions 
while $1-p_b$ is the probability of sampling complex ones.
Similarly, we experiment with two distance measure implementations
for the $d$ function of \Cref{eq:dis} that correspond to the $l^2$ (or Euclidean)
and $l^\infty$ norms:
\begin{align}
\small
l^2(m(B_1), m(B_2) &= \displaystyle \sum_{1 \leq i \leq \mid \mathbb{L} \mid + 1} \sqrt{(m(B_1)^{(i)} - m(B_2)^{(i)}}) \\
l^\infty(m(B_1), m(B_2)) &= \displaystyle \max_{1 \leq i \leq \mid \mathbb{L} \mid + 1} \mid m(B_1)^{(i)} - m(B_2)^{(i)} \mid
\end{align}

For SODGA and MODGA parameters, we turn to established literature values
to determine the population size and further selection details.
However, applications of evolutionary fuzzing to
compiler testing are scarce, and established evolutionary fuzzing
tools like \textsc{VUzzer} \cite{rawat2017vuzzer} and
\textsc{V-Fuzz} \cite{li2019v} make no recommendation in these regards.
As such, we turn to standard values used in search-based
testing literature, particularly of \textsc{EvoSuite}, which uses
a population size of 50 and a \textit{tournament selection} with tournament size of~10
\cite{fraser2011evosuite, panichella2017automated}.
We apply the mutation and recombination (crossover) operators
described in Section~\ref{sec:operators}.

Past research by \citet{arcuri2013parameter}
has shown that default values can provide solid
performance in a broad set of scenarios in software testing.
Though these findings provide no guarantees for the task of compiler
testing in particular, we believe these are sensible
starting points that circumvent the expensive requirements of parameter
tuning of all proposed algorithms.

\subsection{Experimental Protocol}

To understand how the simplicity bias influences the nature
of generated files, we perform several runs of RS with different parameter settings.
We experiment with simplicity bias values between 0.4 and 0.6, as we empirically found
that values outside this range either lead to blocks too large to scale
effectively, or too small to build sufficiently complex programs.
% We run RS for 90 minutes for each setting of simplicity bias,
We collect information regarding the number of Kotlin programs generated,
their size, and the types of compiler crashes each simplicity bias reveals.

To answer RQ2, we run both SODGA and MODGA with $l^2$ and $l^\infty$ norms
and we analyze the properties of the generated programs in a manner analogous 
to that done for RQ1.
Finally, we answer RQ3 by considering the effectiveness and efficiency
of RS, and the GAs with best-performing parameter values (for diversity)
based on the results of the previous experiments (RQs).
We measure effectiveness in terms of the number of differential bugs uncovered by each algorithm 
at the end of each search/fuzz run. For efficiency, we collect the number of 
bugs uncovered over time (every 180 seconds) and compute the  Area-Under-Curve (AUC) of the resulting
bugs-over-time graph. 
For RQ3, we perform statistical analysis using the Wilcoxon \cite{conover1999practical} signed-rank test to determine whether the underpinning distributions are significantly different.

In total, we ran five instances of RS to answer RQ1,
three instances of SODGA and MODGA to address RQ2,
and ten runs of RS and SODGA to answer RQ3.
Each run lasts 90 minutes
for a total of 28 total fuzzing sessions amounting to 42 hours
of fuzzer runtime.
In each run, we store \textit{snapshots} of the population of SODGA
and of the archive of MODGA every 180 seconds.
We carry out all DT procedures on
Kotlin version \texttt{1.8.20-RC-release-288}%
\footnote{https://github.com/JetBrains/kotlin/releases/tag/v1.8.20}, which contains both K1 and K2 releases in the same package\footnote{url{https://kotlinlang.org/docs/whatsnew1820.html}}.

All experiments and runs were executed in isolated containers using \textit{Docker}
and performed on the same machine using an
AMD Ryzen 7 5800H with \SI{16}{\giga\byte} of RAM.

\begin{figure*}[th]
\begin{minipage}[b]{0.4\textwidth}
\centering
\includegraphics[scale=0.4]{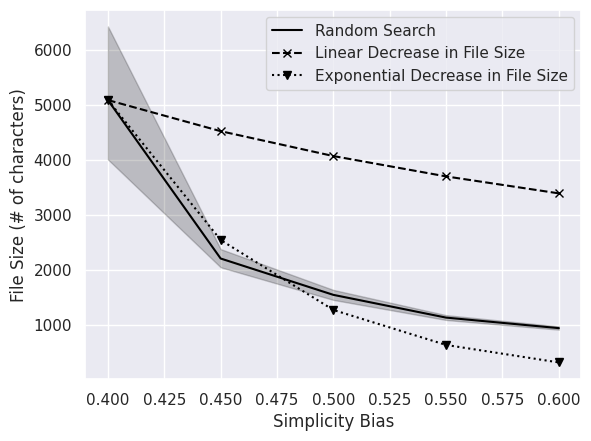}\\
\subcaption{Mean size of a block generated by RS as a function of simplicity bias.}
\end{minipage}%
\hspace{1cm}
\begin{minipage}[b]{0.4\textwidth}
\centering
\includegraphics[scale=0.4]{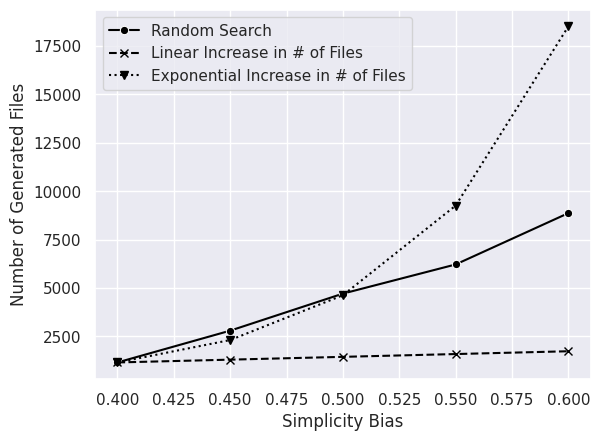}\\
\subcaption{Number of blocks generated by RS per 90 minutes as a function of simplicity bias.}
\end{minipage}%
\caption{Comparison of the number of files generated by RS and their size.}
\label{fig:rq1-size-and-number}
\end{figure*}

%%%%%%%%%%%%%%%%%%%%%%%%%%%%%%%%%%%%%%%%%%%%%%%%%%
\section{Results and Analysis}
\label{sec:results}
%%%%%%%%%%%%%%%%%%%%%%%%%%%%%%%%%%%%%%%%%%%%%%%%%%

%In this section, we detail the results of the empirical study, answering each research question separately.

\subsection{Effects of Expression Simplicity}

To investigate the impact of the simplicity bias, we analyze the size and number of files generated by the Random Search (RS) algorithm when varying the simplicity bias $p_b$.
\Cref{fig:rq1-size-and-number} depicts the relationship between 
(a) the average size of the Kotlin programs generated through random sampling and 
(b) the number of files the algorithm outputs in a 90-minute interval.
To help contextualize the rate of change, we include additional data points representing linear changes (depicted as dashed lines) and exponential changes (depicted as dotted lines) in the data. For (a), each previous value is halved, while for (b), each previous value is doubled.
% We limit the analysis of simplicity bias values to a range between 0.4 and 0.6,
% which we empirically evaluated to strike a balance between complexity and size.
% Values of this parameter exceeding $0.60$ tend to severely
% limit the semantic context in which code is generated,
% while values below $0.40$ increase the probability of producing
% output whose size and complexity diminish its real-world application.

For simplicity bias values between 0.40 and 0.50, both rates of change
are comparable to their exponential counterparts.
The average size of a file decreases from 5,084 characters for a bias of 0.40
to 2,205 for a bias of 0.45 and 1,545 for a bias of 0.50.
The rate of change diminishes for bias values 0.55 and 0.6, with
average sizes of 1,132 and 941 characters, respectively.
Conversely, the number of generated files (programs) increases from 
1,157 for the lowest value of bias (0.40) to 8,869 for the highest (0.60).
Values corresponding to biases between the two extremes follow the same trend,
resulting in 2,799, 4,712, and 6,218 files generated for simplicity biases 
of 0.45, 0.50, and 0.55, respectively. 
Both rates of change consistently and significantly outpace their linear equivalents. 
In the first two intervals, the rate of change also surpasses its exponential counterpart.

\begin{figure}[t]
    \centering
    \includegraphics[width=0.80\linewidth]{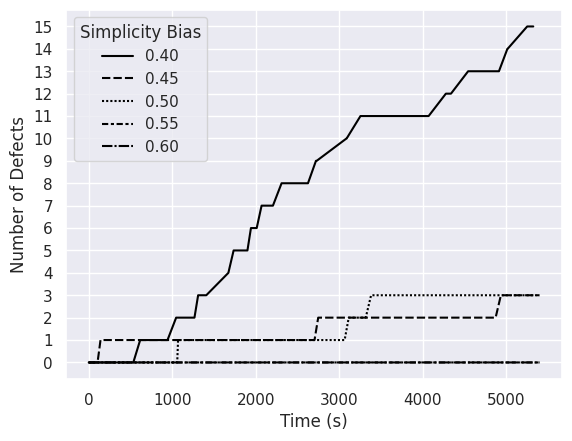}
    \caption{The number of defects uncovered by RS with varying simplicity bias values.}
    \label{fig:rq1-1convergence}
\end{figure}

Next, we analyze the relationship between the simplicity bias
and the effectiveness of random sampling in identifying differential bugs 
between the two Kotlin compilers, namely \ka and \kb.
\Cref{fig:rq1-1convergence} presents a visual representation of the number of 
defects discovered through differential testing of the generated files, 
corresponding to each tested simplicity bias value.
A crucial finding from these results is that \textit{all} defects uncovered by 
RS are of a single type: out-of-memory (OOM) errors, specifically in the 
\texttt{K1} compiler. 
%, and not in its newer \texttt{K2} counterpart.

All OOM errors encountered in the five experimental runs are triggered by files with 10,000+ characters. Their distribution among the
simplicity bias values aligns with the distribution
of file sizes that each value entails.
RS with a simplicity bias of 0.40 generates 15 OOM-inducing
files, while higher values of 0.45 and 0.50 both produce 3 such instances.
Simplicity bias values exceeding 0.50 do not generate such files,
due to their narrower file size distribution.
\\
\sumbox{
\textbf{Summary RQ1}: Lower simplicity bias values cause RS
to generate larger and fewer files.
Bias values of 0.50 or lower occasionally generate files of over 
10,000 characters, which often trigger OOM errors in \ka, but not in \kb.
}

\subsection{Effects of Diversity Interpretation\label{subsec:diversity}}

\begin{figure*}[t]
\begin{minipage}[b]{0.4\textwidth}
\centering
\includegraphics[scale=0.4]{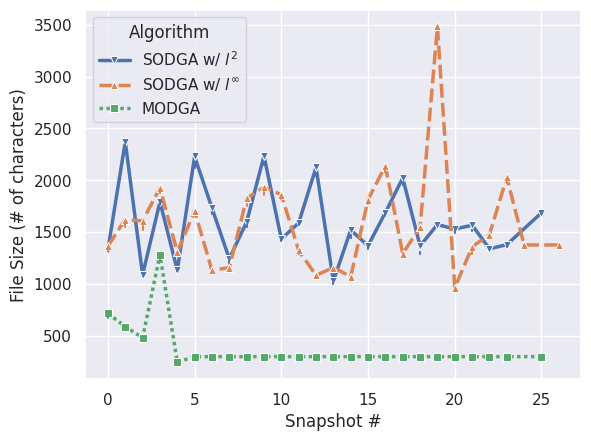}\\
\subcaption{Mean size of blocks generated by SODGA and MODGA.}
\end{minipage}%
\hspace{1cm}
\begin{minipage}[b]{0.4\textwidth}
\centering
\includegraphics[scale=0.4]{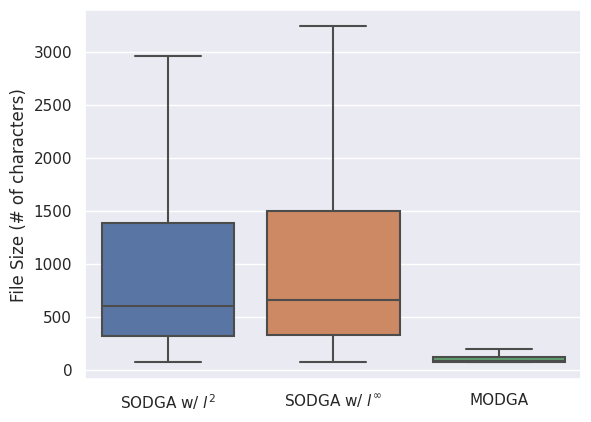}\\
\subcaption{Block size distribution of SODGA and MODGA.}
\end{minipage}%
\caption{Block size distribution of SODGA and MODGA.}
\label{fig:rq2-dist}
\end{figure*}

%%%
% We consider the implications of using the $l^2$ and $l^\infty$ norms as distance
% measures for SODGA.
% We additionally take the behavior of MODGA into account as a point of comparison,
% which does not utilize distance metrics in its optimization.
We begin by analyzing the properties of the Kotlin programs generated and collected in
each snapshot of SODGA and MODGA with $l^2$ and $l^\infty$ norms.
\Cref{fig:rq2-dist} (a) provides the visualization of the evolution of the
average generated file size over time.
For both variations of SODGA, the mean file size
varies between 1,000 and 3,500 characters,
which are well under the threshold of 10,000 characters
that may trigger OOM errors.

The fluctuations stem from the influence of the population-sensitive fitness function.
As the population tends towards larger programs, the diversity-based fitness function (and the selection operator)
promotes smaller programs as they will be more diverse (i.e., more distant regions of the search space).
As smaller programs take over the population, the trend reverses, and larger
programs again introduce more diversity.
This pattern holds true for both the $l^2$ and $l^\infty$ norms.
However, the latter norm induces more substantial shifts in program sizes due to its more rigorous measure of (dis)similarity. Consequently, the $l^\infty$ variant yields both the largest (3,500) and smallest (900) average program sizes.

In contrast, the elitist archive of MODGA retains files that
are far smaller than the population of its single-objective counterparts.
Following initial fluctuations in the first six snapshots, the average program size in the population of MODGA stabilizes at around 300 characters. This aligns with the composition of the archive itself, which also converges to a static set of 70 solutions by the sixth snapshot. Subsequent changes over time are comparatively minimal, with the archive expanding by only three additional entries.

\Cref{fig:rq2-dist} (b) depicts the program size distribution generated by SODGA and MODGA.
The means of the distributions of the two SODGA variations
are comparable: the $l^\infty$ variant produces files that are, on average, only
$1.50\%$ (1,574 characters) smaller than $l^2$ (1,598 characters).
However, the two distributions substantially differ in their third and fourth interquartile, with the $l^\infty$ norm generating more outliers.
% This property reveals itself in the median and IQR measures,
% which are $9.8\%$ and $10.3\%$ (658 and 1177) larger for 
% the $l^\infty$ optimizer than in the Euclidean revision (599 and 1067).
MODGA produces files that are both much smaller and less diverse than either
of the SODGA.
This pattern can be attributed to two primary factors: (1) the size component notably reduces the archive's overall size; 
(2) the archive has a tendency to retain only a few (less than 10) very large files, which independently dominate substantial portions of the search space.

In terms of bug finding, the genetic algorithms allowed to discover two bugs.
SODGA equipped with the $l^\infty$ norm and MODGA
both independently uncover one defect that causes
\texttt{K1} to raise an error, while \texttt{K2} successfully
compiles the generated code.
Both instances of the defect emerge as a consequence of the genetic
recombination operator, and could not have been otherwise generated by RS.
We discuss this further in \Cref{subsec:analysis}. \\
\sumbox{
\textbf{Summary RQ2}: SODGA behaves similarly
when using $l^2$ and $l^\infty$ norms,
with the latter displaying a broader program distribution.
%Both variations fluctuate in terms of average program size,
%based on the algorithm's recent history.
MODGA's archive stagnates after a few iterations and retains smaller files.
All GAs uncover defects that RS could not discover.
}
\vspace{-10pt}

\subsection{Comparative Analysis of the Detected Bugs}

To gain additional insights into the comparative performance of the
two classes of algorithms, we compare the
defect detection capabilities of RS and SODGA.
To this end, we performed 10 fuzzing sessions with RS and SODGA w/ $l^2$.
We selected this variant of single-objective GA since it generates
fewer outlier files than its $l^\infty$ counterpart, and does not suffer
from possibly premature convergence like MODGA.
% We also selected STPGA-200 rather than the \gls{MO}-based counterparts
% because the latter's extreme archive growth would make for an unfair
% comparison to the other two algorithms, that do not retain such a mechanism.
Both algorithms are run with a simplicity bias of $0.5$ based on the results of RQ1.

\begin{figure}[t]
    \begin{tikzpicture}
[scale=0.3,
  circ/.style={shape=circle, inner sep=17pt, draw, node contents=}]
    \draw node (c1) at (-3, 2) [circ, label=below:{\small (RS: 2)}];
    \node[] at (-3,2)   (a) {
    \begin{minipage}{2cm}
    \centering
    \small Resolution \\ Ambiguity \\ \cref{fig:concur}
    \end{minipage}};

    \draw node (c2) at (3, 2) [circ];
    \node[] at (3,2)   (b) {
    \begin{minipage}{2cm}
    \centering
    \small OOM \\ Errors
    \end{minipage}};

    \node[] at (3,-2.25)   (b) {
    \begin{minipage}{2cm}
    \centering
    \small (RS: 10) \\ (SODGA: 1)
    \end{minipage}};

    \draw node (c2) at (9, 2) [circ, label=below:{\small (SODGA: 8)}];
    \node[] at (9,2)   (b) {
    \begin{minipage}{2cm}
    \centering
    \small Conflicting \\ Overloads \\ \cref{fig:nested-funcs}
    \end{minipage}};

    \draw node (c2) at (3, 2) [circ];
    \node[] at (-7,5)   (b) {RS};
    \draw (-8,-3.5) rectangle (6,6);

    \node[] at (12,5)   (b) {SODGA};
    \draw [dashed] (0,-3.5) rectangle (14,6);
    
    \end{tikzpicture}
    \caption{Overview of uncovered defects.}
    \label{fig:venn}
\end{figure}
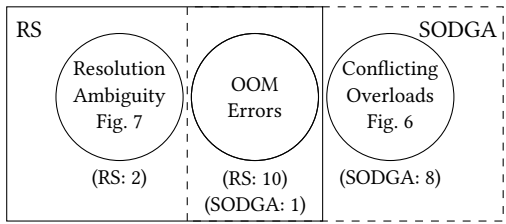

% \Cref{fig:rq3-conv} (a) and (b)
% show the convergence plots emerging from the two experiments.
In total, RS uncovers more defects (12 unique bugs) than SODGA (9 unique bugs).
However, the Wilcoxon test revealed no statistically significant
difference \wrt{} the number of detected bugs between RS and SODGA ($p=0.459$).
The RS runs reveal one novel defect that causes \texttt{K2}
to not compile a generated Kotlin program when \texttt{K1} does.
SODGA additionally finds 1 OOM error and 8 defects of the same category as the ones
addressed in \Cref{subsec:diversity}, but no novel defects emerge. 
RS achieves, on average, a mean AUC of $0.719$ (\ie{} bug detection over time),
which is greater than SODGA's $0.446$. However, the pairwise statistical analysis 
suggests that these differences are not statistically significant ($p=0.322$).
In other words, SODGA and RS show similar efficiency in finding bugs over time.
\Cref{fig:venn} provides an overview of the categories of bugs each
algorithm uncovers. Of the 21 bugs found in total, 11 are OOM errors
that RS finds more frequently (10) than SODGA (1).
RS additionally uncovers 2 instances of a resolution ambiguity defect,
and SODGA distinctly finds 8 instances of conflicting overload errors.
% We discuss these instances in more detail in \Cref{subsec:analysis}.
 \\
\sumbox{
\textbf{Summary RQ3}: 
RS and SODGA with $l^2$ norm are statistically equivalent \wrt{} the number of bugs they uncover and their efficiency (bugs detected over time).
However, the two algorithms are complementary as they find distinct categories of defects.
}
\vspace{-5pt}

\subsection{Bugs Confirmed by JetBrains Developers \label{subsec:analysis}}

Throughout the experiments carried out in this study, we encountered
several dozen instances of erroneous behavior in the Kotlin compiler.
After individual analysis and consultation with the Kotlin compiler
developer team, we divided these instances into three distinct categories.
This section briefly analyzes these categories, their impact on the Kotlin ecosystem,
and the components of the fuzzer that were responsible for their detection.

\subsubsection{OOM Errors}
OOM errors occurring only in the \texttt{K1} compiler are common among files that exceed
10,000 characters in length.
The fact that such files do not cause equivalent errors in \texttt{K2}
makes them less important for the scope of this study, as they showcase a measurable
improvement in performance, rather than an issue that requires developer attention.
As a result, we did not report any of these issues to the developer team,
instead focusing on bugs that affect \texttt{K2}.
We occasionally encountered code that either
(i) triggered OOM errors in both \texttt{K1} and \texttt{K2},
or (ii) only triggered OOM errors in \texttt{K2}.
Compiler developers confirmed several such instances for the current release of Kotlin,
but consider them of minor importance or even acceptable.

\begin{figure}[t]
    \centering
    \begin{lstlisting}
    [
    frame=leftline,
    framesep=2mm,
    baselinestretch=1,
    % bgcolor=gray,
    fontsize=\footnotesize,
    linenos
    ]
    {java}
    fun main() {
        fun p () : Char { return 'c' }    
        fun p () : Float { return 13.0f }
    }
    \end{lstlisting}
    \caption{\texttt{K2} false negative conflicting overloads.}
    \label{fig:nested-funcs}
\end{figure}

\subsubsection{\texttt{K2} Nested Functions}

\Cref{fig:nested-funcs} contains a code snippet that 
\texttt{K2} compiles without warnings,
while \texttt{K1} throws a \textit{conflicting overloads} error.
The latter is the intended behavior.
The two \texttt{p()} functions cause a resolution conflict that the compiler
is meant to warn about.
After experimentation with this instance, we observed that \texttt{K2} always
resolves a \texttt{p()} function call to the first definition of the function,
irrespective of the return type or the number of re-declarations. Notably, this problem only occurs when the definitions of \texttt{p()}
are both \textit{nested} inside a higher-scope function.
The Kotlin compiler developers confirmed the existence of this bug and assigned it a medium priority.
They plan to fix this error in Kotlin 2.0.

We uncovered several independent instances of this bug, all generated through GAs.
The sampling process that RS depends on
contains a check that prevents the generation of functions with
the same name, which is tracked through the shared context.
However, this constraint is relaxed during recombination,
allowing for such scenarios to emerge during crossover occasionally.

\subsubsection{\texttt{K2} Concurrent Modification Exception}

\Cref{fig:concur} shows an error that affects
the \texttt{ConcurrentModificationException} class of the Kotlin standard library.
The \texttt{K2} compiler reports an overload resolution ambiguity error
stemming from the expression in line 4.
\texttt{K1} compiles the code without error.
The Kotlin developers verified the occurrence of this bug in several recent compiler releases.
The compiler team traced the bug to the resolution and inference components
of the compiler frontend, and, after careful consideration, decided this change is expected.
Though any permutation of algorithm and configuration described in this study is in
principle capable of generating this, it was RS that initially uncovered it.
Finding such bugs depends on covering the input
of the fuzzers' context, which RS is more effective at since it does not incur additional
overhead from operations on higher-level language structures.

\begin{figure}[t]
    \centering
    \begin{lstlisting}
    [
    frame=leftline,
    framesep=2mm,
    baselinestretch=1,
    % bgcolor=gray,
    fontsize=\footnotesize,
    linenos
    ]
    {java}
    var J_: ConcurrentModificationException =
            ConcurrentModificationException()
    J_ = J_ ?: ConcurrentModificationException(
            ConcurrentModificationException(J_))
    \end{lstlisting}
    \caption{\kb~overload resolution ambiguity.}
    \label{fig:concur}
\end{figure}

% 2.5 Page
%%%%%%%%%%%%%%%%%%%%%%%%%%%%%%%%%%%%%%%%%%%%%%%%%%
\section{Threats to Validity}
\label{sec:validity}
%%%%%%%%%%%%%%%%%%%%%%%%%%%%%%%%%%%%%%%%%%%%%%%%%%

Threats to \textit{construct validity} stem from the connection between 
the practical measures employed to quantify theoretical aspects of the study.
In this regard, we use standard DT procedures to quantify
compiler defects and link uncovered defects to static, measurable properties of
the generated code.
We also employ common metrics of effectiveness and efficiency that
are standard practices throughout empirical software engineering research.

Threats to \textit{internal validity} regard factors that could lead
to confounding the causes of observed phenomena.
In our study, the main obfuscating factor in regards to causality
is the heavy reliance on randomness.
We take several measures to address the large degree of randomness
inherent to fuzzing approaches.
First, we individually assess each uncovered defect
and analyze the components of the fuzzer involved in its generation.
Through this procedure, we identify the root causes of generating
compiler-crashing code and isolate the merits and disadvantages of different heuristics.
Second, we ensure the fairness of
the comparisons by sharing parameters
between different configurations with
default values from the literature.
We also perform the comparison using a single implementation
of the tool, that relies on the same sampling process and genetic operators.
Lastly, to assess the effectiveness and efficiency of our algorithms,
we repeat independent runs 10 times and report average values
that are subject to standard statistical analyses.

Threats to \textit{conclusion validity} affect the relation between the available data
and the credibility of the conclusions we derive based on it.
To this end, we base our analysis on over 50,000 generated files,
and vary the essential hyperparameters of our algorithms to several
sensible values.
We additionally perform statistical tests to compare algorithms representative
of their respective class.
In particular, we base our performance conclusions on
the application of the Wilcoxon signed-rank test, which
is a statistical procedure that does not
impose unreasonable restrictions on the data distribution.

Threats to \textit{reproducibility} concern factors that might cause the application
of the same research methods to result in significantly divergent or conflicting observations.
To mitigate this, we supply the entire code
base that implements our approach \cite{kotfuzz-impl},
in addition to the entire set of generated
files and adjacent preprocessed data~\cite{kotfuzz-data}.
We also provide extensive documentation
to detail our tool, its configuration, and its applicability.
To ensure that no environmental factors interfere with the fuzzer,
we use containerization to isolate the dependency management and runtime of our tool.
%%%%%%%%%%%%%%%%%%%%%%%%%%%%%%%%%%%%%%%%%%%%%%%%%%
\section{Conclusions and Future Work}
\label{sec:conclusion}
%%%%%%%%%%%%%%%%%%%%%%%%%%%%%%%%%%%%%%%%%%%%%%%%%%

We proposed a generalizable three-stage approach that intertwines
syntax, semantics, and meta-heuristic search.
Our method prunes semantically rich grammar productions
and replaces them with context-aware samplable counterparts
to drastically decrease the likelihood of generating invalid code.
We structure the code that emerges from sampling
the enriched grammar structure into a hierarchical
representation based on scope and complexity.
This representation forms the basis
of an evolutionary framework that provides guidance
to the sampling process.

We introduced two instances of evolutionary algorithms
that are novel to the field of compiler fuzzing.
The algorithms seek to
drive the population toward a diverse collection of code
that exercise different combinations of language features.
We implemented both single- and many-objective
formulations of genetic algorithms (GAs), in addition to
the standard random sampling (RS).

We analyzed the behavior and performance of the proposed
approaches in an empirical analysis spanning 50,000 generated Kotlin files,
which we analyzed through differential testing between the recent \texttt{K1}
compiler and the upcoming \texttt{K2} version.
Our results uncovered three previously unreported categories of bugs,
which we reported to the Kotlin compiler developer team.
The developers verified and replicated our instances
on the current release of the Kotlin compiler.
Compiler developers are either working on fixing the reported issues,
or have already resolved them in more recent compiler releases.
While a comparative analysis between RS and GAs shows no significant difference in the number of bugs they find, their driving mechanisms 
favor different code patterns, which in turn materialized
in distinct bugs uncovered by each heuristic.
We foresee multiple possible directions for future work.

\textit{Further grammar enrichment.} The current version supports custom sample operations for a limited number of Kotlin CFG rules. Extending these to additional rules would allow the fuzzer to generate more varied and interesting code efficiently.

\textit{Compiler Integration}.
The heuristics and fitness
functions explored in this study
 guide the generative sampling process towards
promising areas of the Kotlin code space.
In parallel to those methods, one could leverage compiler information
directly into the search algorithm.
Our tool includes a module that allows the fuzzer
to query for files that trigger faults, according to DT.
%However, the fuzzer does not currently integrate this information into the search process.
%Integration could follow different blueprints, such as
%maintaining code that uncovers faults as a permanent part of the population 
%or avoiding generating novel blocks that resemble known faulty code.
%The former promotes the generation of more instances of the same bug,
%while the latter prevents it.

\textit{Integration with Mutation Fuzzing} \citet{stepanov2021type} introduced
a mutation-based fuzzer for Kotlin, that alters input code
in a sound and type-aware manner.
Since the variation operators of our GAs are both vastly
different and less powerful than those in the mutation fuzzer,
exploring the integration of the two could give rise to new,
otherwise unattainable pieces of code.
%This integration could follow a two-step approach,
%where files generated through our method serve as input to the mutation fuzzer.
%Alternatively, a closer integration of the two methods
%would use the mutation fuzzer as a powerful variation operator as part of a GA.

% 0.5 Page

%\balance

\bibliographystyle{ACM-Reference-Format}
\bibliography{bibliography}

\end{document}